# HISTOGRAM ANALYSIS OF GALLEX, GNO AND SAGE NEUTRINO DATA: FURTHER EVIDENCE FOR VARIABILITY OF THE SOLAR NEUTRINO FLUX

Peter A. Sturrock[1] and Jeffrey D. Scargle[2]

1. Center for Space Science and Astrophysics, Varian 302G, Stanford University, Stanford, CA 94305, USA
2. NASA/Ames Research Center, MS 245-3, Moffett Field, CA 94035

## ABSTRACT

If the solar neutrino flux were constant, as is widely assumed, the histogram of flux measurements would be unimodal. On the other hand, sinusoidal or square-wave modulation (either periodic or stochastic) may lead to a bimodal histogram. We here present evidence that the neutrino flux histogram is in fact bimodal. We analyze all available data from gallium experiments, coordinating results from the GALLEX and GNO experiments into one data set, and adopting results from the SAGE experiment as another data set. The two histograms, from the two data sets, are consistent in showing peaks in the range 45 - 75 SNU and 90 - 120 SNU, with a valley in between. By combining the data into one data set, we may form more detailed histograms; these strengthen the case that the flux is bimodal. A preliminary statistical analysis indicates that the bimodal character of the solar neutrino flux is highly significant. Since the upper peak is close to the expected flux (120 - 140 SNU), we may infer that the neutrino deficit is due to time-varying attenuation of the flux produced in the core. We estimate the time scale of this variation to be in the range 10 – 60 days. Attenuation that varies on such a time scale is suggestive of the influence of solar rotation, and points towards a process involving the solar magnetic field in conjunction with a nonzero neutrino magnetic moment.

## 1. INTRODUCTION

The results of solar neutrino experiments present a number of puzzles that have been reviewed and analyzed by Bahcall and others (Bahcall 1989, Bahcall et al. 1996). These analyses are based on the assumption that the solar neutrino flux is constant. A number of authors have looked for a correlation between the solar neutrino flux and an index of solar variability such as the Wolf sunspot number (Bahcall, Field & Press 1987; Bahcall & Press 1991; Bieber et al. 1990; Dorman & Wolfendale 1991), the surface magnetic field strength (Massetti & Storini 1993; Oakley et al. 1994), the intensity of the green-line corona (Massetti & Storini 1996), and the solar-wind flux (McNutt 1995). Some (but not all) of these authors claim that the studies give positive results and show evidence of variability on a time-scale of years, but these claims have been criticized by Walther (1997). However, there is independent evidence for variability on shorter time-scales. We have found in our analysis of the Homestake data that the scatter is larger than one would expect from a constant flux, and that the neutrino flux varies with a period comparable with that of solar rotation, indicating a dependence of the flux on solar longitude (Sturrock, Walther & Wheatland 1997). We have also found that the flux displays a seasonal variation that may be attributed to a dependence upon solar latitude (Sturrock, Walther & Wheatland 1998). Preliminary analysis of GALLEX data also provides evidence for rotational modulation (Sturrock, Scargle, Walther, & Wheatland, 1999).

The issue of time variability has crucial significance for the resolution of the neutrino problem. If the flux is indeed constant, then the most likely interpretation of the deficit is the MSW effect (Mikhevev & Smirnov 1986a, 1986b, 1986c; Wolfenstein 1978, 1979) as is generally believed, although the VVO effect (Voloshin, Vysotskii, & Okun 1986a, 1986b) and the RSFP effect (Akhmedov 1988a, 1988b; Lim & Marciano 1988) cannot be ruled out. On the other hand, if the solar flux varies on a time-scale of days or weeks, we must go back to square one. We may then consider the possibility that nuclear burning is variable and perhaps is not spherically symmetric, due to some presently unrecognized instability: in this case, the deficit may still be due to the MSW effect, and a combination of asymmetric burning and the MSW effect may lead to modulation at the solar rotation frequency. However, if nuclear burning is constant and is spherically symmetric, some process other than the MSW effect must be involved, since the MSW effect depends only on mass density which, in the Sun, is very close to being spherically symmetric.

If the solar neutrino flux is variable, and if the possibility of fluctuations or asymmetry of the nuclear burning process may be ignored, then one must conclude that the flux is modulated in transit from the core to the Earth, presumably in the radiative zone or in the convection zone. If the variation is periodic, the relevant period will be a clue to the location of the region in which modulation occurs. This modulation must be spatially inhomogeneous, indicating that the mechanism involves the Sun's internal magnetic field and either the VVO effect or the RSFP effect. Either mechanism requires that the neutrino magnetic moment be nonzero.

It is difficult (but, we believe, not impossible) to determine from radiochemical data whether or not the solar neutrino flux varies on a time-scale of days or weeks since the data are acquired in runs, each of which may last from two to six weeks. Data from the Kamiokande and Super-Kamiokande experiments (Fukuda et al. 1998) are not yet available for analysis. In this article, we present another approach to the problem of deciding whether or not the solar neutrino flux is constant. We consider histograms formed from flux estimates obtained from the GALLEX (Anselmann et al. 1993, 1995; Hampel et al. 1996), GNO (Altman et al. 2000), and SAGE (Abdurashitov et al. 1999) experiments. If the neutrino flux were constant, these histograms should have only one peak, i.e. they would be unimodal. On the other hand, if the flux varies in time, there can be more than one peak. In particular, a periodic variation can lead to a bimodal distribution. For instance, if the flux z varies sinuoidally,

$$z = C + A\sin(\boldsymbol{w}t), \qquad (1.1)$$

the distribution function for z is given by

$$f(z)dz = \frac{1}{\boldsymbol{p}} \frac{dz}{\left[A^2 - (z-C)^2\right]^{1/2}} \ for \ |z-C| \le A, \ f(z)dz = 0 \ for \ |z-C| > A. \ , \qquad (1.2)$$

which has two cusps.

## 2. BAYESIAN ANALYSIS

The usual approach to analyzing solar-neutrino data is to represent the measurement from each run by a probability-distribution function (PDF). Experimenters give, for each run, an estimate of the flux z (in SNU), and upper and lower one-sigma error bars. By adding to these estimates assumptions about the form of the PDF (say half-gaussian forms above and below the flux estimate), one can construct a PDF for each run. We really need to consider also PDF's representing various possible forms for the measurements to be expected on the basis of one or more hypotheses. We may take into account both sets of PDF's by a Bayesian procedure (Sturrock 1973). We represent by $P(z|r)$ the probability that the true flux measurement for run r was z where, for simplicity, we consider a discrete sequence of values of z, namely integral values. We represent by $P(z|H_a)$ the probability that we should have measured the flux to be z, based on the hypothesis $H_a$. We need to consider a set of hypotheses, say $a = 1,...,A$, and we need to set a "prior probability" $P(H_a|-)$ for each hypothesis. Then the post-probability for each hypothesis, based on information for run r alone, it given by

$$P(H_a|r) = \left[ \sum_z \frac{P(z|H_a)P(z|r)}{\sum_b P(z|H_b)P(H_b|-)} \right] P(H_a|-) . \tag{2.1}$$

The post-probabilities based on information for all runs is given by

$$P(H_a|r=1-R) = K\left[P(H_a|-)\right]^{-(R-1)} \prod_{r=1}^{R} P(H_a|r) , \tag{2.2}$$

where K is chosen so that the post-probabilities sum to unity.

The usual approach is to form the likelihood function

$$L(z) = \prod_r P(z|r) . \tag{2.3}$$

The value of z for which L is a maximum is taken to be the best estimate of the actual flux, and the width of the peak is taken to give an estimate of the uncertainty in the flux estimate. This procedure is effectively the same as the above Bayesian procedure if we adopt a delta function representation of $P(z|H_a)$:

$$P(z|H(Z)) = d_{z,Z} . \tag{2.4}$$

This is a "zero-entropy" or "minimum-ignorance" hypothesis: the other extreme is the "maximum entropy" or "maximum-ignorance" hypothesis H0, for which

$$P(z|H0) = const. \tag{2.5}$$

over the specified range for z, that we take to extend from a minimum value $z_l$ to a maximum value $z_u$. The constant is to be chosen so that $P(z \mid H0)$ sums to unity.

We have considered data for the 84 runs so far reported for the GALLEX and GNO experiments. For each run, we have formed a gaussian distribution P(z|r) centered on the reported flux estimate, and (to be conservative) with standard deviation given by the larger of the lower and upper error estimates. In order to cover all values, negative as well as positive, we adopted $z_l = -200$ and $z_u = 400$. With this choice of data, hypotheses, and parameters, we find that the post probability $P(H(Z) \mid r = 1-84)$ is a maximum for Z = 56. However, this maximum value is of order $10^{-200}$. Compared with the maximum-entropy hypothesis (for which the post-probability is $1 - 10^{-200}$), the delta-function hypothesis fares very badly! If we consider only runs for which the flux estimate is non-negative, we find that the maximum occurs near $z = 71$, which is closer to the estimates cited in Altman et al. (2000).

The ignorance hypothesis does not have any physical significance, since it assigns non-zero prior probability to negative values of the flux. We can form another hypothesis H0n, which is physically significant, by assigning constant prior probability to all non-negative values of z up to the maximum $z_u$. This corresponds to a completely erratic neutrino flux that is equally likely, for any run, to have any value whatever in the prescribed range. If we repeat the above analysis in such a way as to compare H(Z) and H0n, we find that the maximum value of $P(H(Z) \mid r = 1-84)$ is of order $10^{-171}$. The usual zero-entropy or delta-function hypothesis again fares poorly.

### 3. EVIDENCE FOR BIMODAL FLUX DISTRIBUTIONS

We have examined flux estimates obtained by the GALLEX-GNO experiment, and find that the distribution appears to be bimodal. However, the data obtained from any one experiment may be subject to some unrecognized systematic effect that could influence the histogram. For this reason, we begin by comparing the histogram formed from GALLEX-GNO data with that formed from SAGE data.

Published data from the GALLEX-GNO experiment now comprise 84 runs. Published data from the SAGE experiment now comprise 57 runs for which flux estimates are derived from K-line data, and 31 runs for which flux estimates are derived from L-line data. In order to concentrate on the main features of the histogram, we display only the part of the histogram in the range $0 < z < 300$ (77 flux estimates from GALLEX-GNO and 70 from SAGE). In order to have at least 10 counts in at least one bin for each data set, we choose displays with 20 bins. Figure 1(a) shows the histogram formed from the GALLEX-GNO data, Figure 1(b) shows the histogram formed from the SAGE data, and Figure 1(c) shows the histogram formed from the combined data. We see that Figures 1 (a) and (b) both show peaks at 45 – 75 and at 90 – 120, and a valley at 75 – 90. This pattern is clearer in the combined data shown in Figure 1(c) than in either Figure 1(a) or Figure 1(b). It appears therefore, from this comparison, that the solar neutrino flux distribution is bimodal. In Figure 2 (a), (b) and (c), we show higher-resolution histograms for the combined GALLEX-GNO and SAGE data, with 30, 45 and 60 bins, respectively. We see that, as the resolution increases, the evidence for bimodality becomes stronger.

## 4. FURTHER ANALYSIS OF THE FLUX HISTOGRAM

In this section, we make a preliminary assessment of the statistical significance of the apparent bimodal structure of the solar neutrino flux histogram. In order to cover all values of the flux estimate, we adopt $z_l = -100, z_u = 500$. We apply the same Bayesian procedure as in Section 2, with the difference that we now represent the PDF's for the data $P(z | r)$ by delta functions. The functional form of the probability distribution function for each hypothesis will be expressed as f(z) for integer values of z in the range $z_l$ to $z_u$. It is understood that the distribution is to be normalized so that its integral is unity. Hence we may fix the height of any part of the curve arbitrarily, and we have chosen $f = 100$ for $z < 0$.

With these conventions, the ignorance hypothesis H0, previously introduced in Section 2, is specified by

$$f = 100 \; for \; z_l \leq z \leq z_u. \tag{4.1}$$

The unimodal hypothesis H1 is here represented by

$$f = 100 \text{ for } z_l \leq z < z_{1a},$$
$$f = f_{1a} \text{ for } z_{1a} \leq z < z_{1b}, \quad (4.2)$$
$$f = f_{1b} \text{ for } z_{1b} \leq z \leq z_u.$$

This model then has the four adjustable parameters $z_{1a}, z_{1b}, f_{1a}, f_{1b}$. The bimodal hypothesis H2 is represented by

$$f = 100 \text{ for } z_l \leq z < z_{2a},$$
$$f = f_{2a} \text{ for } z_{2a} \leq z < z_{2b},$$
$$f = f_{2b} \text{ for } z_{2b} \leq z < z_{2c}, \quad (4.3)$$
$$f = f_{2c} \text{ for } z_{2c} \leq z < z_{2d},$$
$$f = f_{2d} \text{ for } z_{2d} \leq z \leq z_u.$$

This model has eight adjustable parameters $z_{2a}, z_{2b}, z_{2c}, z_{2d}, f_{2a}, f_{2b}, f_{2c}, f_{2d}$.

We have evaluated these three hypotheses by the Bayesian method used in Section 2. We find that the post-probability of H1 is a maximum for the following values of the parameters: $z_{1a} = 17$, $z_{1b} = 155$, $f_{1a} = 900$, $f_{1b} = 24$. We assign the same values to the corresponding parameters in H2: $z_{2a} = 17$, $z_{2d} = 155$, $f_{2a} = 900$, $f_{2c} = 900$, $f_{2d} = 24$. Then we find that the ratio of the post-probability of H2 over that for H1 is a maximum for the following values of the remaining three parameters: $z_{2b} = 71$, $z_{2c} = 94$, $f_{2b} = 300$. For this choice of the adjustable parameters, we find that $P(H0 \mid D) = 1.2\ 10^{-51}$, $P(H1 \mid D) = 2.9\ 10^{-10}$, and $P(H2 \mid D) \approx 1$. We see that, in contrast with the situation in Section 2, the ignorance hypothesis now fares very poorly. We also see that the bimodal hypothesis is favored over the unimodal hypothesis by an odds ratio of approximately $3\ 10^9$.

The above odds value cannot be interpreted directly as the odds that the distribution of the solar neutrino flux measurements is bimodal rather than unimodal, since we have adjusted several parameters to maximize the post probabilities. Nevertheless, it seems unlikely that the flexibility in our choice of parameters is sufficiently great to reverse the weight of evidence in favor of H2 over H1. Suppose for instance, as a worst-case scenario, that we consider 100 values of each parameter, and that, for each parameter, the post-probabilities are comparable with the above values for only one choice out of 100.

Since H2 has three more adjustable parameters than H1, this would reduce the odds in favor of H2 over H1 by a factor of $10^6$. Even in this overly pessimistic case, the odds in favor of the bimodal hypothesis are of order 3000.

We have also evaluated the optimum parameters for H2, independently of the values derived for H1. In this case, the optimum choice is found to be $z_{2a} = 17$, $z_{2b} = 71$, $z_{2c} = 94$, $z_{2d} = 155$, $f_{2a} = 900$, $f_{2b} = 310$, $f_{2c} = 440$, $f_{2d} = 17$. From these values of the parameters for H2 and the previous values of the parameters for H1, we obtain the following post-probabilities: $P(H0|D) = 5.2\ 10^{-52}$, $P(H1|D) = 1.3\ 10^{-14}$, and $P(H2|D) \approx 1$. This appears to strengthen the case for bimodality, but H2 now has one further adjustable parameter. A more thorough analysis of this question requires us to consider the post probability for a range of values of each parameter, and then to integrate over those parameters with weighting functions corresponding to the prior probabilities of the parameters. We plan to present the results of this calculation in a later article.

## 5. DISCUSSION

Unless there is some previously unrecognized systematic effect that leads to the same bimodal distribution for both the GALLEX-GNO and the SAGE experiments, it is difficult to avoid the conclusion that the bimodal structure of the histograms is due to variability of the solar neutrino flux. As we mentioned in Section 1, there is already independent evidence that the flux is indeed variable. Hence any attempt to understand the bimodal structure of the histograms should be coupled with an attempt to understand the nature and cause of the variability. This variability may be periodic but one should not rule out the possibility that it is stochastic.

Although the timing of runs does not enter explicitly into our calculations, we may infer that the time scale of the variation is not short compared with the half-life of the $^{71}$Ge products (11.43 days), since if it were the variation would be washed out, and the histogram would be unimodal. We have also evaluated the histogram of a time series in which each element is the mean of the flux measurements of two consecutive runs, and find that the resulting histogram is almost unimodal. We may infer from this fact that the time scale of variation is not long compared with the typical run duration (about 30 days). These considerations imply that the variation responsible for the bimodal structure has a time scale in the range 10 to 60 days, which encompasses the range of periods of internal solar rotation (Schou et al. 1999). We have found that when we correct for the rotational modulation of the GALLEX-GNO flux measurements, the histogram becomes approximately unimodal.

We may infer, from the location of the peaks in the histogram (approximately 65 SNU and 105 SNU) that the minimum to maximum ratio of the variation related to the bimodal histogram of flux measurements is approximately 0.6. This corresponds to a "depth of modulation" of approximately 25%. Since the flux measurements correspond to weighted integrals of the $^{71}$Ge production rate, we should expect that the actual depth of modulation of the solar neutrino flux will be found to exceed 25%. The upper shoulder of the histogram (90 – 120 SNU) is indistinguishable from the expected range (120 – 140 SNU; Hampel et al. 1999, Kirsten 1995). Hence the upper limit of measured values may correspond to the actual flux emitted by the solar core.

The search for an explanation of the bimodal nature of the solar neutrino flux distribution, coupled with an analysis of the variability of that flux, will require reconsideration of a number of assumptions concerning the solar interior, as well as reconsideration of some assumptions concerning neutrinos. At this stage of our research, it seems most likely that the variability is due to the interplay of a nonzero neutrino magnetic moment with the Sun's internal magnetic field, due either to the VVO effect or to the RSFP effect. Hence this line of research, if fruitful, may lead to information concerning the neutrino magnetic moment and possibly also the neutrino mass or masses.

## ACKNOWLEDGEMENTS

We wish to acknowledge support (for PAS) by NASA grants NAS 8-37334 and NAGW-2265 and NSF grant ATM-9910215 and (for JDS) by the NASA Applied Information Systems Research Program. We wish to acknowledge helpful discussions with Guenther

Walther and Mark Weber. We are grateful to the referee, Arnold Wolfendale, for constructive criticism of an earlier version of this Letter.

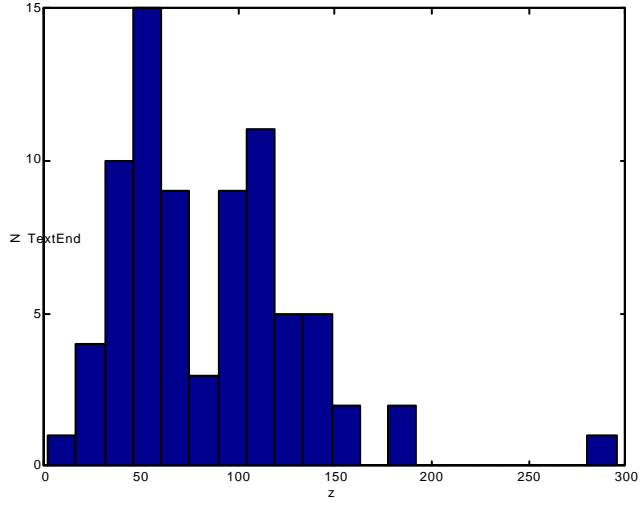
Figure 1a

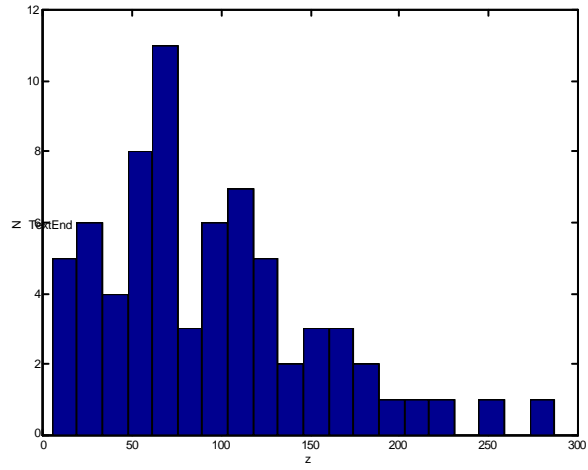
Figure 1b

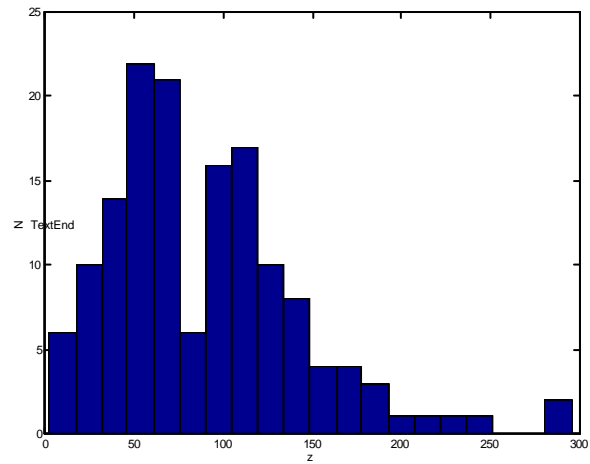
Figure 1c

Figure 1 (a), (b) and (c): Histograms for GALLEX-GNO, SAGE, and combined GALLEX-GNO and SAGE data, respectively, all with 20-bin resolution.

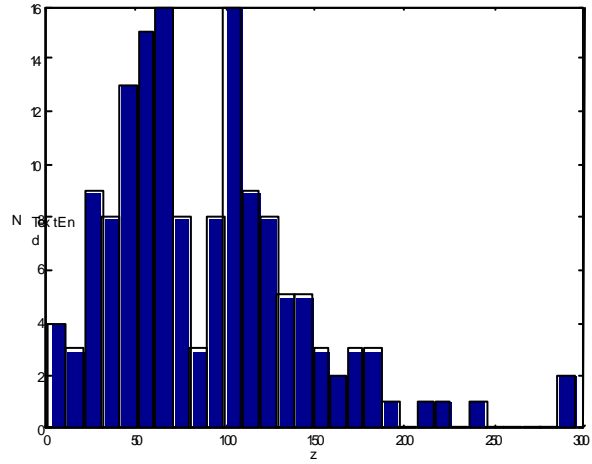

Figure 2a

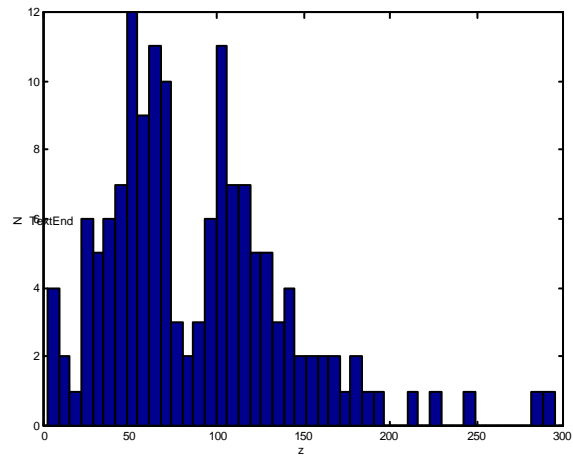

Figure 2b

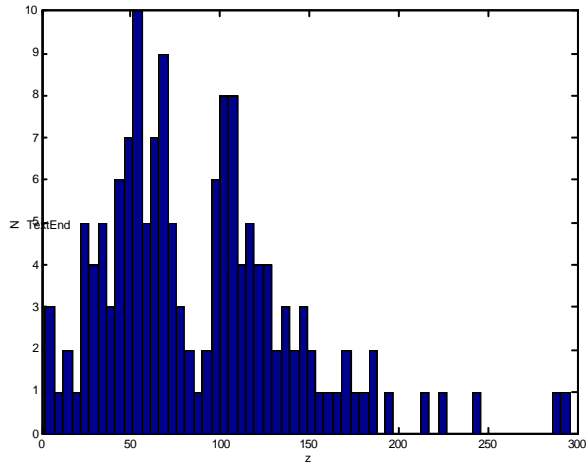

Figure 2c

Figure 2 (a), (b) and (c): Histograms for combined GALLEX-GNO and SAGE data with 30, 45 and 60-bin resolution, respectively.